\documentclass[twocolumn, tighten, astrosymb, a4]{aastex631}

\usepackage{amsmath}
\usepackage{gensymb}

\begin{document}

\title{The Ejection of Transient Jets in Swift J1727.8-1613 Revealed by Time-Dependent Visibility Modelling}

\author[0000-0002-2758-0864]{Callan M. Wood}
\affiliation{International Centre for Radio Astronomy Research, Curtin University, GPO Box U1987, Perth, WA 6845, Australia}

\author[0000-0003-3124-2814]{James C. A. Miller-Jones}
\affiliation{International Centre for Radio Astronomy Research, Curtin University, GPO Box U1987, Perth, WA 6845, Australia}

\author[0000-0003-2506-6041]{Arash Bahramian}
\affiliation{International Centre for Radio Astronomy Research, Curtin University, GPO Box U1987, Perth, WA 6845, Australia}

\author[0000-0002-8195-7562]{Steven J. Tingay}
\affiliation{International Centre for Radio Astronomy Research, Curtin University, GPO Box U1987, Perth, WA 6845, Australia}

\author[0000-0002-8032-7024]{He-Xin Liu}
\affiliation{Key Laboratory for Particle Astrophysics, Institute of High Energy Physics, Chinese Academy of Sciences, 19B Yuquan Road, Beijing 100049, People’s Republic of China}

\author[0000-0002-3422-0074]{Diego Altamirano}
\affiliation{School of Physics and Astronomy, University of Southampton, University Road, Southampton SO17 1BJ, UK}

\author{Rob Fender}
\affiliation{Astrophysics, Department of Physics, University of Oxford, Keble Road, Oxford, OX1 3RH, UK}

\author{Elmar K\"{o}rding}
\affiliation{Department of Astrophysics/IMAPP, University of Nijmegen , 6500 HC Nijmegen, The Netherlands }

\author[0000-0003-1897-6872]{Dipankar Maitra}
\affiliation{Department of Physics and Astronomy, Wheaton College, Norton, MA 02766, USA}

\author[0000-0001-9564-0876]{Sera Markoff}
\affiliation{Anton Pannekoek Institute for Astronomy, University of Amsterdam, Science Park 904, 1098 XH Amsterdam, The Netherlands}
\affiliation{Gravitation and Astroparticle Physics Amsterdam Institute, University of Amsterdam, Science Park 904, 1098 XH 195 196 Amsterdam, The Netherlands}

\author[0000-0002-3500-631X]{David M. Russell}
\affiliation{Center for Astrophysics and Space Science (CASS), New York University Abu Dhabi, P.O. Box 129188, Abu Dhabi, UAE}

\author[0000-0002-7930-2276]{Thomas D. Russell}
\affiliation{INAF, Istituto di Astrofisica Spaziale e Fisica Cosmica, Via U. La Malfa 153, I-90146 Palermo, Italy}

\author[0000-0003-0167-0981]{Craig L. Sarazin}
\affiliation{Department of Astronomy, University of Virginia, 530 McCormick Road, Charlottesville, VA 22904-4325, USA}

\author[0000-0001-6682-916X]{Gregory R. Sivakoff}
\affiliation{Department of Physics, University of Alberta, CCIS 4-181, Edmonton AB T6G 2E1, Canada}

\author[0000-0002-4622-796X]{Roberto Soria}
\affiliation{INAF - Osservatorio Astrofisico di Torino, Strada Osservatorio 20, 10025 Pino Torinese, Italy}
\affiliation{College of Astronomy and Space Sciences, University of the Chinese Academy of Sciences, Beijing 100049, People's Republic of China}
\affiliation{Sydney Institute for Astronomy, School of Physics A28, The University of Sydney, Sydney, NSW 2006, Australia}

\author[0000-0003-3906-4354]{Alexandra J. Tetarenko}
\affiliation{Department of Physics and Astronomy, University of Lethbridge, Lethbridge, Alberta, T1K 3M4, Canada}

\author{Valeriu Tudose}
\affiliation{Institute of Space Science - INFLPR Subsidiary, 077125 Magurele, Romania}



\begin{abstract}
\noindent High angular resolution radio observations of relativistic jets are necessary to understand the causal connection between accretion and jet ejection in low mass X-ray binaries. Images from these observations can be difficult to reconstruct due to the rapid intra-observational motion and variability of transient jets. We have developed a time-dependent visibility model fitting and self-calibration procedure and applied it to a single four-hour VLBA observation of the low-mass X-ray binary Swift J1727.8-1613 during the bright flaring period of its 2023 outburst. This allowed us to detect and model a slightly resolved self-absorbed compact core, as well as three downstream transient jet knots. We were able to precisely measure the proper motion and flux density variability of these three jet knots, as well as (for the first time) their intra-observational expansion. Using simultaneous multi-frequency data, we were also able to measure the spectral index of the furthest downstream jet knot, and the core, as well as the frequency-dependent core shift between 2.3 and 8.3\,GHz. Using these measurements, we inferred the ejection dates of the three jet knots, including one to within $\pm40$ minutes, which is one of the most precise ever measured. The ejection of the transient jet knots coincided with a bright X-ray flare and a drastic change in the X-ray spectral and timing properties as seen by HXMT, which is the clearest association ever seen between the launching of transient relativistic jets in an X-ray binary and a sudden change in the X-ray properties of the accretion inflow.
\end{abstract}



\section{Introduction} \label{sec:intro}
    The causal connection between the accretion inflow and the launching of relativistic jets by black holes is best studied in nearby Galactic black hole low-mass X-ray binaries (LMXBs). During bright outbursts of these systems, the properties of the accretion inflow and the jet outflows change dramatically \citep[see e.g.][for a review of disk/jet coupling in LMXB outbursts]{2004MNRAS.355.1105F, 2009MNRAS.396.1370F}.
    
    To date, there have been multiple suggestions of X-ray intensity, spectral, and timing signatures that are associated with the launching of relativistic jets. One such signature is the switch from type-C to type-B quasi-periodic oscillations \citep[QPOs; for a review, see][]{2019NewAR..8501524I}, which have been observed close to the launch time of transient relativistic jets in a few sources \citep{2012MNRAS.421..468M, 2019ApJ...883..198R, 2020ApJ...891L..29H, 2021MNRAS.505.3393W}. Despite this, the exact causal relationship between the changes in the inner accretion flow and the ejection of the transient jets has not been determined. Our understanding is limited primarily by the lack of precise measurements of the ejection times of jet knots, accompanied by sufficiently dense contemporaneous X-ray observations. While the jet knots can often be tracked out to arcsecond-scale distances \citep[e.g.][]{1994Natur.371...46M, 2002Sci...298..196C}, milli-arcsecond (mas) resolution observations using very long baseline interferometry (VLBI) are essential for precise ejection date measurements \citep[e.g.][]{2012MNRAS.421..468M, 2021MNRAS.505.3393W}. 

    Early VLBI observations of X-ray binary jets were often contaminated by imaging artefacts or had to be reduced in length due to the rapid intra-observational variability of the jet knots \citep[][]{1995Natur.375..464H,1995Natur.374..141T}. This rapid variability violates the fundamental assumption of source stability in aperture synthesis. One way of accounting for this variability is by splitting the full observations into short time-bins where the jet knots are less variable but the sensitivity is lower. This approach has been used to track the launching and evolution of transient jets in VLBI observations; e.g. in Scorpius X-1 \citep{2001ApJ...558..283F}, and V404 Cygni \citep{2019Natur.569..374M}. This approach requires the jet knots to be bright so that they can be detected within each time bin, and imaged with the subsequently sparse $uv$-coverage. Because of the challenges of reconstructing long observations of highly variable jet knots, shorter observations are generally taken over multiple days, with the knots moving between the observations \cite[e.g.][]{1987Natur.328..309V, 2000ApJ...543..373D, 2001ESASP.459..291H}. However, in some cases individual knots have only been detected in a single observation \citep[e.g.][]{2005ASPC..340..281M}. The short time-scale variability of these transient jets is therefore not often studied in high angular resolution observations. 
    
    \cite{2010MNRAS.409L..64Y} devised a method to measure the proper motion of a jet knot from a single observation of XTE J1752-223, by shifting the phase centre of an observation within an image to correct for a given proper motion, varying the proper motion to maximise the recovered flux density of the jet knot. \cite{2021MNRAS.505.3393W} used a similar proper motion correction method to detect a fast-moving knot in MAXI J1820+070 that had been smeared below the noise level of an observation. These approaches do not account for flux density variability or expansion of jet knots, and are not well suited for observations of multiple jet knots moving at different speeds. 
    
    \cite{2023MNRAS.522...70W} introduced a new time-dependent visibility model fitting approach that can account for both large proper motion and flux density variability to measure the properties of the time-evolving jet knots. This technique allows for the precise measurement of time-varying jet parameters from a single high angular resolution observation.

    \subsection{Swift J1727.8-1613}
    Swift J1727.8-1613 was first detected by \textit{Swift}/BAT as it went into a bright outburst beginning in August 2023 \citep{2023GCN.34537....1P}. It was then quickly followed up at X-ray, radio, and optical wavelengths \citep[see e.g.][Hughes et al. in prep.]{2023ApJ...958L..16V, 2024ApJ...971L...9W, 2024ApJ...968...76I, 2024A&A...682L...1M, 2024ApJ...960L..17P}. Swift~J1727.8-1613 was later dynamically confirmed as a black hole LMXB \citep{2025A&A...693A.129M}. VLBI observations early in the outburst showed that Swift J1727.8-1613 had a resolved continuous jet aligned in the north-south direction \citep{2024ApJ...971L...9W}. X-ray observations of the source revealed that after the initial rise of the outburst, the source entered a bright flaring period, which included the brightest X-ray flare of the outburst on 2023 September 19 \citep[see e.g.][]{2024arXiv240603834L, 2024ApJ...970L..33Y, 2024MNRAS.529.4624Y, 2024ApJ...974..303Z, 2024arXiv241006574L}. During this period, rapid changes in the X-ray intensity, spectral, and timing properties were observed, as well as multiple radio flares \citep{2024ApJ...968...76I}. 
    
    In this letter we make use of a time-dependent visibility modelling approach, with newly-implemented self-calibration and multi-frequency modelling procedures, to study a high angular resolution VLBI observation of Swift J1727.8-1613 taken during the bright flaring period.

\section{Observations and Calibration}

    \subsection{VLBA}

        Following the beginning of the rise of the bright X-ray flare on 2023 September 19, we triggered an observation with the VLBA (observation code BM538B), as part of our ongoing monitoring of this outburst \citep[see][]{2024ApJ...971L...9W}. We observed from 2023 September 19 23:56 to 2023 September 20 03:41 (UTC) with the dichroic S/X feed with a recording rate of 4096 Mbps. The data were split into four 128-MHz intermediate frequency (IF) pairs, with the lowest IF pair centered at 2.3\,GHz (S-band) and the other three IF pairs centered at 8.3\,GHz (X-band). We split the data into the two separate bands and calibrated them individually. We used ICRF J174358.8-035004 (J1743-0350) as a fringe finder, ICRF J172134.6-162855 (J1721-1628) as a phase reference source, and ICRF J172446.9-144359 (J1724-2914) as a check source \citep{2020A&A...644A.159C}. 
        
        The data were correlated using the DiFX software correlator \citep{2007PASP..119..318D, 2011PASP..123..275D}, and calibrated according to the standard procedures within the Astronomical Image Processing System \citep[\textsc{aips}, version 31DEC22;][]{1985daa..conf..195W, 2003ASSL..285..109G}. The 2.3\,GHz data were largely corrupted by interference, and so we could only reliably calibrate the final $\sim1$ hour of the observation. Following the standard external gain calibration, we performed several rounds of hybrid mapping of the phase reference source to derive the time-varying phase, delay, and rate solutions, which we interpolated to the target. We also performed a single round of amplitude self-calibration to get the most accurate time-varying amplitude gain calibration, which we then applied to the target.

    \subsection{HXMT}
    
        We analysed a series of observations of Swift~J1727.8-1613 taken by the Hard X-ray Modulation Telescope \citep[HXMT;][]{2014SPIE.9144E..21Z} around the time of the bright X-ray flare on 2023 September 19. These observations are a subset of the data published in \cite{2024MNRAS.529.4624Y} and the data were processed following standard procedures. We filtered the X-ray counts from the low-energy (LE), mid-energy (ME), and high-energy (HE) instruments into three bands covering 2-10 keV, 10-35 keV, 27-250 keV, respectively. We binned the X-ray counts into 10-second averaged light curves for each band using \texttt{stingray} \citep{2019ApJ...881...39H, bachettiStingrayFastModern2024}. We also computed the fractional-rms normalised dynamic power spectra for each good time interval in each band using a 15-second window.

\section{Analysis}
    \subsection{Imaging}

\begin{figure}
            \centering
            \includegraphics[width=\linewidth]{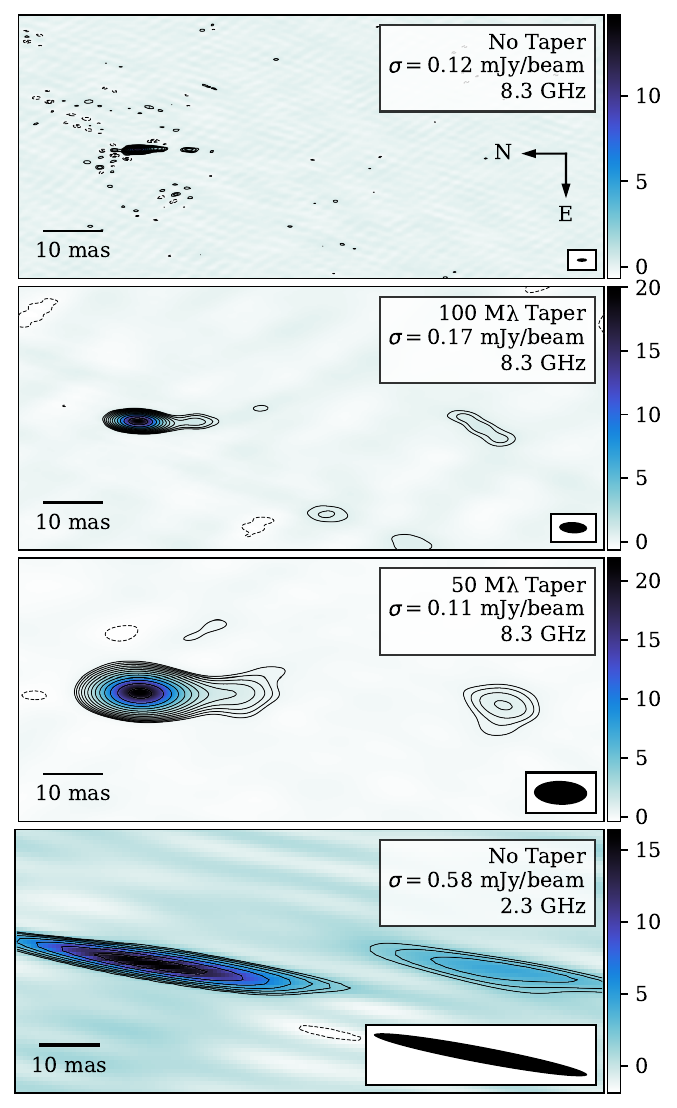}
            \caption{VLBA images of the jets launched by Swift~J1727.8-1613 on 2023 September 19. The first three panels show images of the 8.3\,GHz observations with Gaussian tapers of increasing severity (as noted in the text boxes). The images are rotated counter-clockwise by $90\degree$, as shown by the compass in the top panel. The colour bar shows the intensity in mJy/beam, and the contour levels are $\pm\sigma\times\sqrt{2}^n$ mJy/beam where $n=3,4,5,...$, and $\sigma$ is the rms noise shown in the upper right of each image. The restoring beams of the images are shown by the ellipses in the bottom right of the panels. The images show a compact core and multiple downstream jet knots. The resolved jet knots are large and diffuse and are only detected with the shortest baselines. The bottom panel shows the simultaneous 2.3\,GHz observation, which has lower sensitivity and poorer resolution. The distinct inner jet knots seen at 8.3\,GHz are blended with the core at 2.3\,GHz, but the downstream jet knot is clearly distinct and optically thin.}
            \label{fig:BM538B jet images}
        \end{figure}

        We imaged the calibrated 8.3\,GHz data in DIFMAP, using several rounds of CLEANing, model fitting, and self-calibration. The initial image showed a bright and slightly-resolved core, with a single distinct jet knot $\sim$8 mas to the south, shown in the first panel of Figure~\ref{fig:BM538B jet images}. In order to detect any diffuse or partially resolved-out emission, we made two images using Gaussian $uv$-tapers with 30\% power at 100 and 50\,M$\lambda$ (mega-wavelengths), respectively (with the full $uv$-coverage extending out to $\sim220\,M\lambda$). We show these images in the second and third panels of Figure~\ref{fig:BM538B jet images}. In the tapered images, the core appears to be more extended, and a southern downstream jet knot is clearly detected at a separation of $\sim60$ mas from the core.  

        We similarly imaged the final $\sim1$ hour of the 2.3\,GHz data within DIFMAP, which is also shown in Figure~\ref{fig:BM538B jet images}. The resolution is much lower than that of the 8.3\,GHz data. The image consists of a bright core with an extended structure to the south, likely the blending of multiple distinct components as in the 8.3\,GHz images, with a single expanded southern jet knot $\sim60$ mas downstream. Since the restoring beams of the 2.3 and 8.3\,GHz observations are so different, it is difficult to estimate the spectral index of the inner blended components. By fitting the downstream jet knot with an extended circular Gaussian in both bands, we measured a steep apparent spectral index of $\alpha=-2.0\pm0.2$ ($S_\nu\propto\nu^{\alpha}$), where we assume a 10\% calibration error for the 8.3\,GHz flux density and 20\% for the 2.3\,GHz flux density (due to the RFI contaminated system temperature measurements). 

        We split the 8.3\,GHz observation into two halves and imaged each of them separately to search for evidence of variability. We found that between the two images, the downstream jet knot appeared to move $\sim5$ mas. This could explain the steep spectral index, due to the reduction in surface brightness owing to the motion smearing the emission across multiple beams in the 8.3-GHz image. The spectral index may have also been steepened due to the shorter spacings of the 2.3 GHz data which may detect resolved emission that we cannot detect in the 8.3 GHz data, although we expect motion to be the dominant factor. We explore this further in section \ref{sec:multi-frequency results}. We also found evidence of motion of the inner diffuse knot at a separation of $\sim$15 mas from the 8.3-GHz core. This inner knot also appeared to expand and start becoming resolved between the first and second images. In both bands, the bright core was stationary, suggesting that it was a compact, steady jet close to the location of the central black hole. 

    \subsection{Time-Dependent Model Fitting}
        Although we could image the two halves of the 8.3\,GHz observation, the sensitivity and $uv$-coverage were insufficient to reliably detect the downstream knots in multiple short time bins. To study the intra-observation motion and variability of the jet components, we used the time-dependent model fitting procedure first described in \cite{2023MNRAS.522...70W} and used again in \cite{2024ApJ...971L...9W}. In this procedure, we fit simple analytical model components (e.g. point sources, Gaussians) directly to the measured visibilities, explicitly parameterising the variability of the component positions, sizes, and flux densities. 

         We model the central compact jet using a stationary elliptical Gaussian component with major and minor full-width half maximum sizes $\Phi_{\text{major}}$ and $\Phi_{\text{minor}}$, respectively, and position angle $\theta$. Similar to \cite{2023MNRAS.522...70W} and \cite{2024ApJ...971L...9W}, we fit the flux density evolution of all of the jet knots using a linear model with flux density $F_0$ at the reference time $t_0$ (which is chosen as 2023-09-20 02:00 UTC), and flux density derivative $\dot F$. For the discrete moving jet knots, we use circular Gaussian components with a ballistic motion in a radial coordinate system. We force the moving knots to move away from the fit location of the compact core along a position angle $\theta$ with speed $\mu$, with a separation $r_0$ at the reference time. We model the FWHM size of the components with both a constant model and a linear model, where $\Phi_0$ is the size at the reference time and $\dot\Phi$ is the expansion rate.

         To perform the model fitting, we use the Bayesian inference algorithm Nested Sampling \citep{10.1214/06-BA127}, implemented in the \texttt{dynesty}\footnote{\url{https://github.com/joshspeagle/dynesty}} Python package \citep{2020MNRAS.493.3132S}. We assume that the noise on the visibilities is Gaussian, and so we use a Gaussian likelihood. We use uniform priors for all parameters, allowing for a suitable range of values based on constraints from the images, except for the position angle along which the knots move and the orientation of the elliptical Gaussian core. For these parameters we use a normal prior with a mean of $180\degree$ East of North and a standard deviation of $2\degree$ \citep{2024ApJ...971L...9W}.
        
        \subsubsection{Single-Frequency Modelling and Self-Calibration}

            We began our modelling with the 8.3\,GHz data. In the absence of calibration errors, \cite{2023MNRAS.522...70W} demonstrated with simulations that it was possible to reliably solve for the motion and variability parameters directly from the visibilities. Since there were still significant residual phase calibration errors in our data after external gain calibration and phase referencing, we needed to incorporate self-calibration into the dynamical model-fitting procedure. Similar to DIFMAP and AIPS, we implemented a procedure that involves several rounds of iterative model fitting and self-calibration to mitigate the residual calibration errors and solve for the time-variable model parameters. We could not implement this approach in \cite{2023MNRAS.522...70W} since the observations did not have sufficient signal-to-noise to derive reliable self-calibration solutions.
    
            We began by fitting a stationary elliptical Gaussian component with linear flux density evolution to the bright core. We initially flagged all baselines to NL and SC, since the external phase gain calibration was particularly poor for those stations. Once the fit converged, we subtracted this model from the data. The residual image showed faint remaining core emission, and a peak $\sim70$ mas downstream, consistent with the southern knot in the third panel of Figure~\ref{fig:BM538B jet images}. 
            
            To perform self-calibration, we used the median of the posterior probability distribution to compute a time-resolved model of the observed visibilities. We then used this model to perform phase only self-calibration, where we solved for the phase of the time-dependent station-based gains on a 1-minute solution interval (including for NL and SC). We then applied these solutions to the externally gain-calibrated observation, and performed a second round of model fitting on this self-calibrated observation with the same elliptical Gaussian core and a moving circular Gaussian component at the location of the southern discrete jet knot. 
    
            We performed this same loop twice more, first inspecting the residual image, performing phase-only self-calibration and then re-fitting the model with an additional moving circular Gaussian component at the location of an intensity peak in the residual image, until the residual image was noise-like with no clear structure. Finally, once all of the moving jet knots were included in the model, we allowed the size of the moving knots to change linearly with time. We then performed a final round of phase-only self calibration, followed by model fitting and a round of phase and amplitude self-calibration with a 30-minute solution interval. We then perform the final model fit to the interatively self-calibrated data.

        \subsubsection{Single-Frequency Results}

\begin{figure*}
                \centering
                \includegraphics[width=\linewidth]{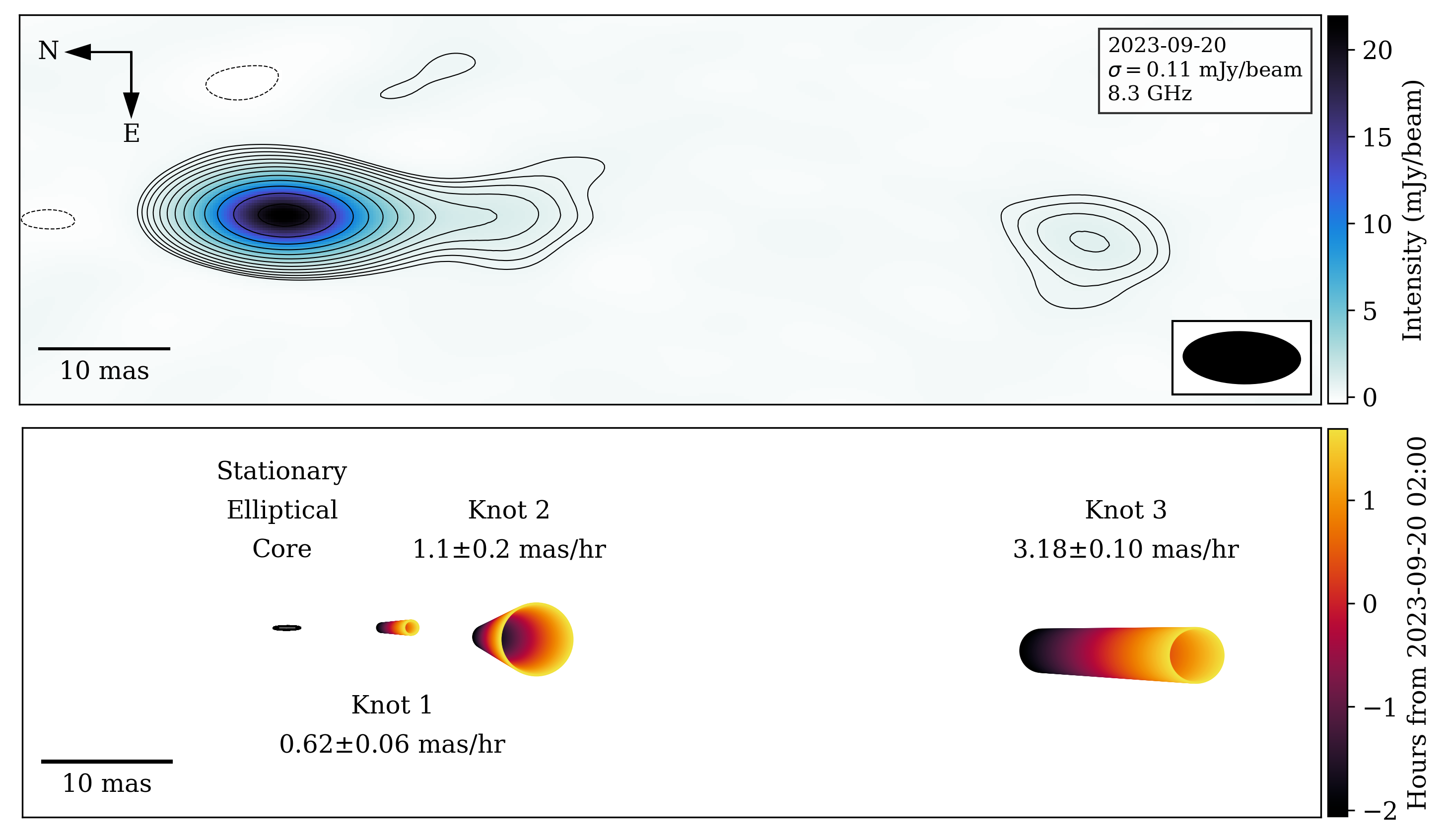}
                \caption{Comparison of our analysis of the VLBA observation of Swift J1727.8-1613 using conventional imaging techniques and our dynamic model-fitting approach. The top panel shows the same image of Swift J1727.8-1613 as in the third panel of Figure~\ref{fig:BM538B jet images}, with the same rotation, colour scale, and contours. The bottom panel shows a representation of our dynamic model fit, on the same scale. We plot the model using the median of the marginal posterior distribution for each fit parameter. The positions and sizes of the ellipses and circles show the evolution of the positions and FWHM sizes of the Gaussian components fit to the visibilities, where the colours of the components correspond to the time in hours from the reference time. The core component is shown in black since it is stationary and not changing in size. The downstream jet knots are moving away from the elliptical Gaussian core at varying speeds, and are all expanding at different rates. An animated version of this figure is available as a supplementary figure.}
                \label{fig:BM538B Image and Median Model}
            \end{figure*}

            In total, we found that the observation was best represented with a stationary elliptical core and three moving circular Gaussian components to the south of the core, which we label knot 1, knot 2, and knot 3. We show a visualisation of this model in Figure~\ref{fig:BM538B Image and Median Model}, where we plot the time-varying locations and sizes of the jet knots from the median of the posterior probability distribution. We include an animated version of this figure as supplementary material. In Table~\ref{tab:single frequency fit parameters} we summarise the parameters fitted in the 8.3\,GHz model. 

\begin{figure*}
                \centering
                \includegraphics[width=\linewidth]{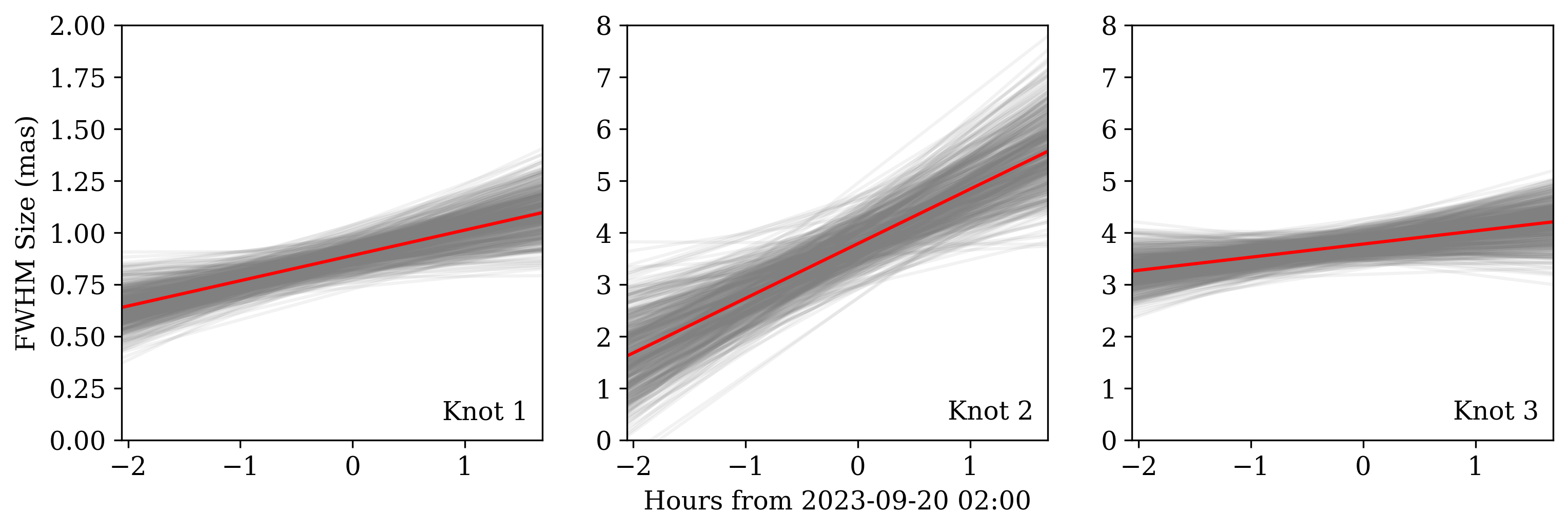}
                \caption{The fitted sizes of the downstream jet knots in our dynamic model of our 8.3\,GHz VLBA observation of Swift~J1727.8-1613. The red lines show the medians of the marginal posterior distributions, while the grey lines are 100 random samples from the posterior distribution. The three jet knots are different sizes and are apparently expanding at different rates.}
                \label{fig:BM538B Jet Size}
            \end{figure*}

            We find that the three jet knots are travelling at different angular speeds, with knot 1 being the slowest and knot 3 being the fastest. The three knots are also all expanding during the observation at apparently different rates, as shown in Figure~\ref{fig:BM538B Jet Size}. We find that the core and knot 3 are decreasing in flux density throughout the observation, while the light curve of knot 1 is consistent with being flat, and knot 2 is increasing in flux density as it expands and moves away from the core. Note that this measurement is of the integrated flux density of the resolved components, and not the peak intensity. 

            \begin{deluxetable}{lll}
                \label{tab:single frequency fit parameters}
                \tabletypesize{\scriptsize}
                \tablewidth{0pt} 
                \tablecaption{Summary of the posterior estimates for the moving jet knots in the VLBA observation shown in Figure~\ref{fig:BM538B Image and Median Model}. The reference time is defined as 02:00:00 (UTC) on 2023 September 20th, which is approximately the midpoint of the observation. The core position coordinates, $x_0$ and $y_0$ (in the directions of R.A. and Dec., respectively), are given relative to the phase centre of the observation (R.A. (J2000) $=17^{\text{h}} 27^{\text{m}} 43^{\text{s}}.31358$, Dec. (J2000) $=-16\degree 12^\prime 19^{\prime\prime}.181$). We report the medians of the marginal posterior distributions as the best-fit parameters, and the 16th and 84th percentiles as the uncertainties.}
                \tablehead{
                \colhead{Component} & \colhead{Parameter}& \colhead{Posterior Estimate}
                }
                \startdata
                Core & $\theta$ (\degree East of North) & $180.41\pm0.10$             \\
                     & $\Phi_{\text{major}}$ (mas)      & $2.015\pm0.012$             \\
                     & $\Phi_{\text{minor}}$ (mas)      & $0.268\pm0.006$             \\
                     & $x_0$ (mas)                      & $0.110\pm0.001$             \\
                     & $y_0$ (mas)                      & $0.153\pm0.004$             \\
                     & $F_0$ (mJy)                      & $20.39\pm0.05$              \\
                     & $\dot F$ (mJy/hr)                & $-2.30_{-0.05}^{+0.04}$     \\
                \hline
                Knot 1 & $\Phi_0$ (mas)                  & $0.89\pm0.05$               \\
                      & $\dot\Phi$ (mas/hr)              & $0.12\pm0.04$               \\
                      & $r_0$ (mas)                      & $8.54_{-0.08}^{+0.07}$      \\
                      & $\mu$ (mas/hr)                   & $0.62\pm0.06$               \\
                      & $\theta$ (\degree East of North) & $180.13\pm0.16$             \\
                      & $F_0$ (mJy)                      & $1.72\pm0.06$               \\
                      & $\dot F$ (mJy/hr)                & $0.00\pm0.06$               \\
                \hline
                Knot 2 & $\Phi_0$ (mas)                  & $3.8\pm0.4$                 \\
                      & $\dot\Phi$ (mas/hr)              & $1.0\pm0.3$                 \\
                      & $r_0$ (mas)                      & $17.4\pm0.2$                \\
                      & $\mu$ (mas/hr)                   & $1.1\pm0.2$                 \\
                      & $\theta$ (\degree East of North) & $177.4\pm0.4$               \\
                      & $F_0$ (mJy)                      & $1.55\pm0.10$               \\
                      & $\dot F$ (mJy/hr)                & $0.36\pm0.08$               \\
                \hline
                Knot 3 & $\Phi_0$ (mas)                  & $3.78\pm0.18$               \\
                      & $\dot\Phi$ (mas/hr)              & $0.25_{-0.17}^{+0.16}$      \\
                      & $r_0$ (mas)                      & $64.66\pm0.11$              \\
                      & $\mu$ (mas/hr)                   & $3.18\pm0.10$               \\
                      & $\theta$ (\degree East of North) & $178.27\pm0.05$             \\
                      & $F_0$ (mJy)                      & $3.03_{-0.09}^{+0.10}$      \\
                      & $\dot F$ (mJy/hr)                & $-0.70\pm0.09$              \\
                \hline
                \enddata
                \tablecomments{The position angle $\theta$ for the core component is the position angle of the major axis of the elliptical Gaussian, where for the three circular Gaussian jet components it is the position angle of ballistic motion away from the core.}
                \tablecomments{The errors listed in this table are the purely statistical errors of the fit to the self-calibrated data. While the relative positions and motions of the components are precisely measured, the absolute positions of the components should contain an additional $98\mu$as and $330\mu$as astrometric error in R.A. and Dec., respectively  \citep{2006AA...452.1099P}. Similarly, the absolute flux densities and flux density derivatives should include a 10\% absolute amplitude calibration error for comparison with other instruments/observations.}
            \end{deluxetable}
            
        \subsubsection{Incorporating the 2.3\,GHz Data}
            Following the single-frequency 8.3\,GHz modelling, we sought to include the 2.3\,GHz data in a combined multi-frequency, time-variable model. We followed the same basic procedure as with the single-frequency data, beginning with a single core component, iteratively performing phase-only self-calibration, and adding more components until we had an elliptical Gaussian core component and three moving circular Gaussian components. In the multi-frequency model, each component had the same position and motion at both frequencies, but they were allowed to have different sizes and flux densities at each frequency. Because the core was optically thick, the position of the core component was shifted downstream at lower frequencies. We therefore allowed the core position to shift with frequency along the jet axis, but forced the positions of the optically-thin jet knots to be the same at both frequencies. This essentially used the optically thin downstream knots to `lock' the reference frame between the two frequencies. 

        \subsubsection{Multi-frequency Results}\label{sec:multi-frequency results}

            In the multi-frequency model, we found that the motions of the moving jet knots were consistent with the single-frequency model. The location, size, and flux density of the core component at 8.3\,GHz was also consistent with the single-frequency model. We found that the elliptical Gaussian core was unresolved at 2.3\,GHz, with upper limits (84th percentile) on the major and minor full-width half maximum sizes of $0.8$ mas and $0.13$ mas respectively. It is not clear why the core component is more compact at 2.3\,GHz than at 8.3\;GHz (see Table~\ref{tab:single frequency fit parameters}). We were able to measure the shift of the centroid position of the core component between 2.3 and 8.3\,GHz to be $3.78\pm0.15$ mas to the south along the position angle of the jet axis. We found that the core component had a spectral index of $0.1\pm0.2$, and the furthest downstream knot (knot 3), had a spectral index of $-1.3\pm0.2$, again assuming a 10\% calibration error for the 8.3\,GHz flux density and 20\% for the 2.3\,GHz flux density, added in quadrature with the statistical errors. This validates the assumption that the downstream jet knot is optically thin, and therefore our core shift measurement is reliable, since we are effectively measuring the difference in separation between this component and the centroid position of the bright, compact core. 
            
            The size, expansion, and 8.3\,GHz flux density of knot 3 was consistent between the single-frequency and multi-frequency modelling. The size, expansion, and 8.3\,GHz flux density of the inner two discrete jet knots showed differences between the single and multi-frequency modelling. This could be due to the poor external calibration and resolution of the 2.3\,GHz data, which resulted in us not being able to properly resolve the inner two components at the lower frequency; or it could be because there was more diffuse emission near the core, and thus the simple two circular Gaussian component model does not adequately describe the jet structure at 2.3\,GHz; or it could be a combination of both. We do not therefore report spectral indices for these two inner knots, and we only use the single-frequency 8.3\,GHz modelling results to determine the proper motions and inferred ejection dates of the knots.

            To test if the spectral indices were artificially steepened by the mismatched uv-coverage of the two frequency bands, we re-performed the multi-frequency time-resolved modelling with the shortest baselines of the 2.3 GHz data removed to match the shortest baselines of the 8.3 GHz data. We found no significant difference in the measured spectral indices with the altered uv-coverage.

        \subsubsection{Transient Jet Ejection Dates}

            Since we were able to model the compact core of Swift~J1727.8-1613, we do not require accurate astrometry to infer the ejection dates of the three jet knots, since we could use the location of the core component to define a fixed reference frame that is conserved through phase self-calibration. Assuming that the core shift evolves with frequency as $\Delta r\propto\nu^{-1}$ \citep{1979ApJ...232...34B}, then by using the measured core shift between 2.3 and 8.3\,GHz we calculate that the 8.3\,GHz core is $1.45\pm0.06$ mas downstream from the central black hole. Using the motions and locations of the transient jet knots from the single-frequency 8.3\,GHz modelling, we infer the ejection dates of knots 1, 2, and 3 to be MJD $60206.41_{-0.07}^{+0.06}$, $60206.36_{-0.17}^{+0.11}$, and $60206.22\pm0.03$, respectively. These measurements are some of the most precisely measured ejection dates of transient jet knots in LMXBs, with the most precise measurement (knot 3) having an uncertainty of $\pm40$ minutes.

        \subsection{X-ray Light Curves and Power Spectra}

\begin{figure}
            \centering
            \includegraphics[width=\linewidth]{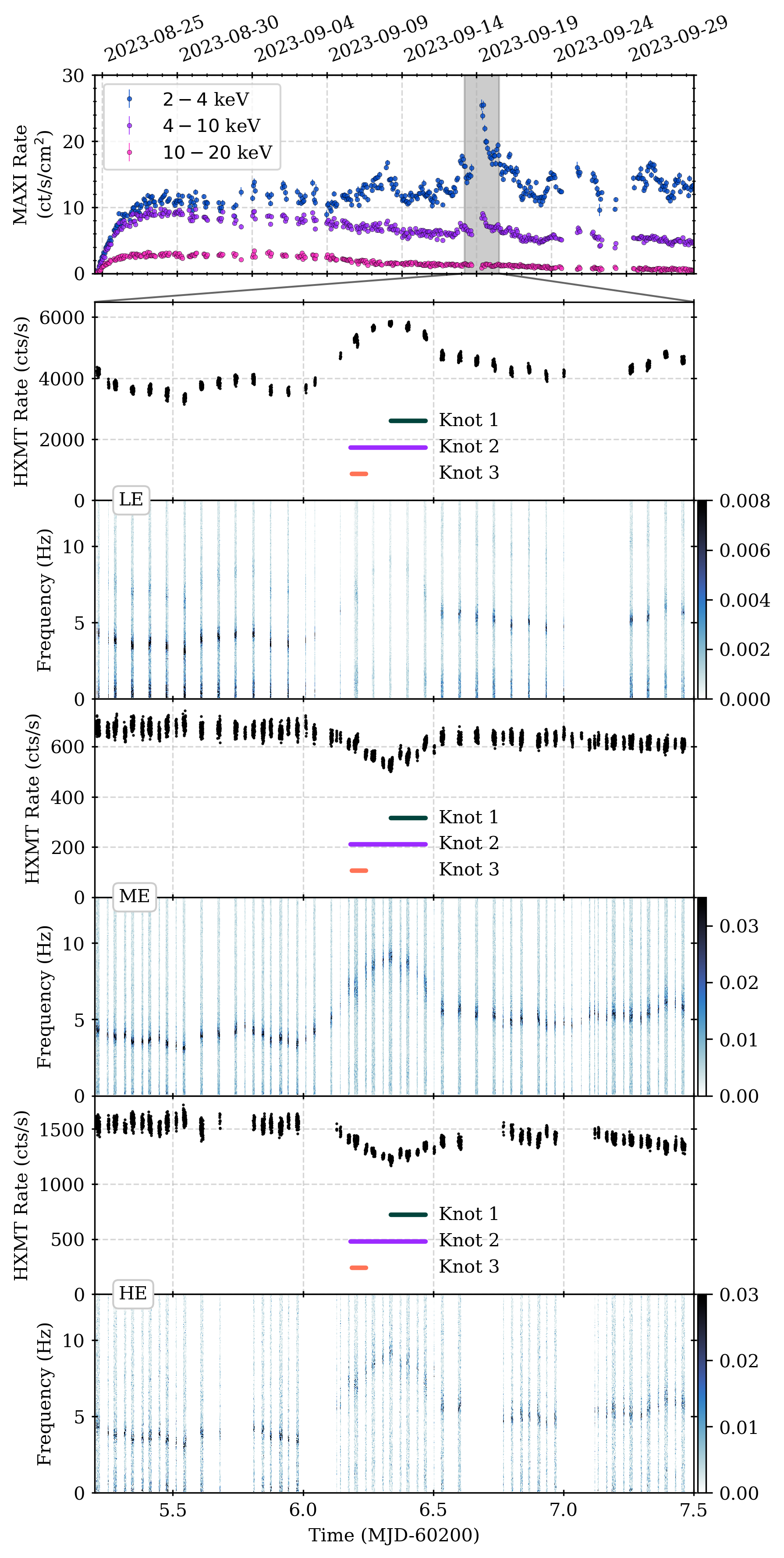}
            \caption{MAXI/GSC light curves of Swift~J1727.8-1613 during the beginning and flaring period of the outburst, with light curves and dynamic power spectra from HXMT in the LE (2-10 keV), ME (10-35 keV), and HE (27-250 keV) bands, surrounding the bright flare on 2023 September 19 (MJD 60206). The colour scale of the dynamic power spectra show the fractional rms normalised power. We note that we have adjusted the colour scale to cut off the maximum power to highlight the fainter features. The horizontal lines show the 16th and 84th percentile credible interval estimations of the ejection dates of the three jet knots shown in Figure~\ref{fig:BM538B Image and Median Model}, corrected for the distance between the measured core position and the central black hole, estimated from the 2.3 to 8.3\,GHz core shift. The ejection of the three jet knots is coincident with the bright X-ray flare and change in spectral and timing properties.}
            \label{fig:lc and dps}
        \end{figure}

        In Figure~\ref{fig:lc and dps} we plot the light curves and dynamic power spectra of the HXMT observations surrounding the inferred ejection dates of the three downstream jet knots. For context, we also include longer-duration light curves from the Monitor of All-sky X-ray Image \citep[MAXI\footnote{http://maxi.riken.jp/};][]{2009PASJ...61..999M} Gas Slit Camera (GSC). The shaded region in the first panel shows the time range of the lower six panels. The ejection dates of the three transient jet knots are all coincident with the bright X-ray flare in the LE band on 2023 September 19 (MJD 60206). At the same time, the ME and HE bands show a dip in their light curves, with the overall X-ray spectrum softening during this flare. In all three bands there is a strong QPO with a centroid frequency around $\sim4$ Hz, which increases to $\sim9$ Hz during the flare, and then recovers back to around $\sim5$ Hz after the flare, following a similar shape to the LE light curve during the flare.

\section{Discussion}
    We made use of our new modelling technique, which, for the first time, included an iterative self-calibration approach, to measure the motions, flux density variability, and expansion of multiple jet knots within a single observation. This is the first time intra-observational expansion has been measured in an LMXB jet knot. We were also able to jointly model two frequency bands to measure the shift in the position of the optically thick stationary core. With traditional imaging alone we could not constrain any of these parameters. 

    \subsection{Jet Properties}

    The brightest component in the 8.3\,GHz observation was the extended, stationary core. We found that this component was optically thick, with a flat or inverted spectral index and a substantial core shift between 2.3 and 8.3\,GHz, which is consistent with being a compact self-absorbed continuous jet. The continuous jet is much less resolved in this observation than in earlier observations in the hard-intermediate state \citep{2024ApJ...971L...9W}. Since it is only marginally resolved, its position is likely close to the location of the approaching jet photosphere at each frequency (i.e.\ the receding jet contribution is very small). The frequency dependent core-shift was first measured in a Galactic X-ray binary in SS 433 \citep{1999A&A...348..910P}, but it has also been measured in the LMXBs V404 Cygni and MAXI J1820+070 \citep{2017ApJ...834..104P, 2021MNRAS.504.3862T, 2023MNRAS.525.4426P}. Our core shift measurement is much larger than the core shifts measured in V404 Cygni and MAXI J1820+070, even accounting for the differences in distance and radio luminosity, although these measurements were derived from observations taken in the hard state.To infer the distance between the measured core and the jet base, we assumed a conical geometry with no free-free absorption in the absence of strong stellar or disk winds. \citet{2020MNRAS.495.3576K} show that the switch from parabolic to conical geometries in nearby AGN occurs at distances of $10^5-10^6$ gravitational radii, whereas our core shift measurement is on the order of $10^8$ gravitational radii (assuming a $10M_\odot$ central black hole), vindicating our assumption of a conical geometry on the scales we observed.

    {We assume that the profile of the core component is Gaussian for the purposes of our modelling, but we note that this likely an oversimplification of the compact jet profile \citep[see, e.g., the compact jet profile in][]{2024ApJ...971L...9W}. While more physically-motivated jet profiles have been proposed \citep[e.g.][Zdziarski et al. in prep.]{2006ApJ...636..316H, 2013MNRAS.432.1319P} an exploration of these models in the visibility plane, and their underlying assumptions, is beyond the scope of this work.
    
    The orientation of the core, the direction of the core shift, and the direction of the proper motion of the jet knots are consistent with the position angle of the extended continuous jet seen at the beginning of the outburst \citep{2024ApJ...971L...9W}. Given that we only detected transient jet knots to the south of the stationary core, that the core shift was to the south, and that the jet was not observed to undergo any large scale precession \citep[e.g.][]{2019Natur.569..374M}, the moving jet components were likely all approaching us. Since we did not detect any receding counter-parts to the jet knots, and since we do not know the inclination of the jet axis, we cannot uniquely constrain the intrinsic speed of the jet knots. The apparent proper motion of an approaching jet knot, $\mu$, is related to the intrinsic speed of the jet knot, $\beta$, by
    \begin{equation}\label{eqn:proper motion}
        \mu=\frac{\beta\sin i}{1-\beta\cos i}\frac{c}{d},
    \end{equation}
    where $i$ is the inclination angle of the jet to the line of sight, $d$ is the distance to the source, and $c$ is the speed of light \citep{1999ARA&A..37..409M}. We assume a distance of $3.7\pm0.3$ kpc \citep{2025A&A...693A.129M}, however we note that different assumed distances \citep[e.g.][]{2025arXiv250206448B} will result in different jet speed, expansion, and inclination constraints (Zdziarski et al. in prep.). We can use the measured proper motion to calculate possible combinations of $\beta$ and $i$ for the individual jet knots. In Figure~\ref{fig:jet speeds} we show the allowed combinations of $i$ and $\beta$ for the moving jet knots. This plot shows that all three jet components must be travelling at different intrinsic speeds, or that they must have drastically different inclination angles. A large difference in their inclination angles is unlikely given that we don't observe a large difference in their position angles. Knot 3, the fastest moving jet, is highly superluminal with an apparent speed of $1.63\pm0.05$ c. This sets an upper limit on the inclination of the jet axis of $i\lesssim67 \degree$, which corresponds to an intrinsic speed $\beta=1$. The inclination upper limit, combined with the dynamical mass function, sets a lower limit on the mass of the black hole in Swift J1727.8-1613 of $3.55\pm0.12 \;M_\odot$ \citep{2025A&A...693A.129M}. Using this inclination upper limit, the minimum intrinsic speeds of knots 1 and 2 are $\beta>0.3$ and $\beta>0.5$, respectively. Knot 3 has a minimum intrinsic speed of $\beta>0.83$, which corresponds to a minimum bulk Lorentz factor of $\Gamma>1.8$. 
    
    \begin{figure}
        \centering
        \includegraphics[width=\linewidth]{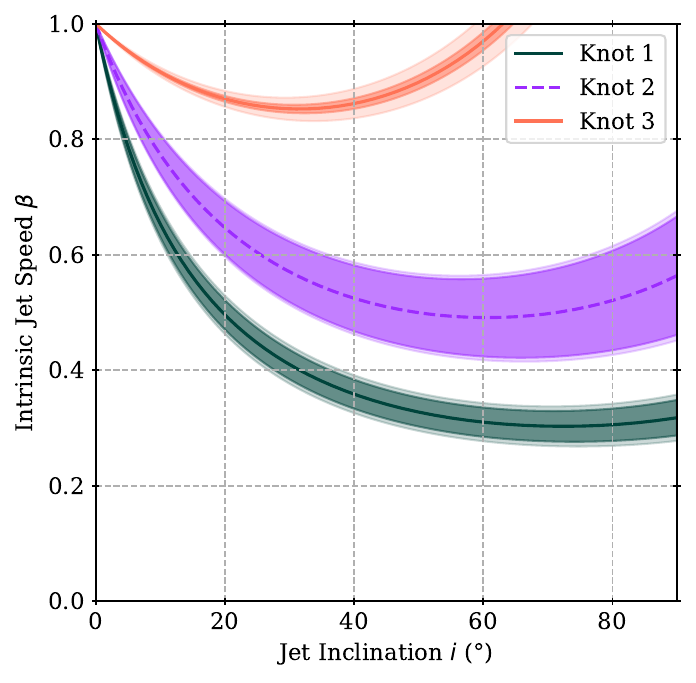}
        \caption{Constraints on the intrinsic jet speed and inclination for the three jet knots shown in Figure~\ref{fig:BM538B Image and Median Model}, using equation \ref{eqn:proper motion}, assuming a distance of $3.7\pm0.3$ kpc \citep{2025A&A...693A.129M}. The lines use the median of the proper motion posteriors, the dark shaded regions include only the uncertainty in proper motion, while the light shaded regions also include the uncertainty in distance. This does not imply that knots 1 and 2 can be travelling at the same speed at the same inclination, since the knots must be at the same distance.}
        \label{fig:jet speeds}
    \end{figure}

    Similarly, we cannot measure the true expansion rate of the jet knots without knowing the intrinsic jet speed and the inclination of the jet axis. The measured expansion rate, $\dot\Phi$, is related to the true expansion rate, $\beta_{\text{exp}}c$, by,
    \begin{equation}\label{eqn:expansion}
        \dot\Phi = \frac{\beta_{\text{exp}}}{\Gamma(1-\beta\cos i)}\frac{c}{d},
    \end{equation}
    \citep{2004ApJ...603L..21M}. So, for the allowed combinations of $\beta$ and $i$ constrained by the proper motion, we can calculate the intrinsic jet expansion rate. We find that for inclinations under the $65\degree$ upper limit, the intrinsic expansion speed of knots 1 and 3 are consistent with each other within uncertainty, although this is dominated by the uncertainty in the apparent expansion speed of knot 3. For all inclinations below $\sim55\degree$, the intrinsic expansion speed of knot 2 is larger than that of knots 1 and 3. Between $55-65\degree$, the intrinsic expansion rate of knot 3 is unconstrained and is consistent with the intrinsic expansion rate of both knots 1 and 3. Knot 1 has the most well constrained intrinsic expansion rate, with a strict upper limit of $\beta_{\text{exp}}<0.08$, and it is inconsistent with the intrinsic expansion rate of knot 2 for all $\beta>0$. At the $65\degree$ upper limit, knot 2 can be expanding as fast as $\beta_{\text{exp}}=0.45\pm0.15$, which is consistent with the relativistic sound speed in plasma ($\beta_{\text{exp}}=1/\sqrt{3}\approx0.577$), although the uncertainty on the expansion speed of this component is quite large.

    It is not clear why the three jet knots are travelling and expanding at different intrinsic speeds. The opening angles (and thus expansion rates) have only been constrained in a small number of systems \citep[see e.g.][]{2001MNRAS.327.1273S, 2006MNRAS.367.1432M, 2017MNRAS.472..141M,  2017MNRAS.469.3141T, 2020ApJ...895L..31E, 2021PASA...38...45C, 2021MNRAS.504.3862T, 2023MNRAS.522...70W}. This observation is the first time that the expansion rate of multiple jet knots has been measured within a single observation. Assuming that the expansion rates of the jet knots are constant, we can infer a `zero-size date', which we show alongside their ejection dates in Table~\ref{tab:ejection dates}. For knot 3, we find that its zero-size date is consistent with its ejection date, whereas the other two knots have zero-size dates much later than their ejection dates. The difference between the zero-size date and the ejection date for knot 2 is the most drastic. This violates the assumption that transient jet knots expand adiabatically with constant opening angle, which many simple transient jet models assume \citep[e.g.][]{1966Natur.211.1131V}. While the interstellar medium environment around the source close to the jet axis is likely dynamic, knots 1 and 2 are close to each other and thus were travelling through a similar environment, and so the processes that caused knot 2 to have expanded much faster than knot 1 must be internal to the jet knots. \cite{2024ApJ...967L...7Z} proposed internal composition as an explanation for the differences in properties and propagation distances of compact and transient jet knots. Since the three jet knots were ejected at a similar time, their difference in propagation speed and expansion rates could be the result of different internal composition or internal magnetic pressures. A difference in collimation of the jet knots may be related to their formation mechanism \citep[e.g.][]{2023ApJ...954L..30S}. While we fit simple Gaussian models to the jet knots, this is likely not a good description of the underlying dynamics of the individual jet knots. 

    \begin{deluxetable}{lll}
        \label{tab:ejection dates}
        \tablewidth{0pt} 
        \tablecaption{Ejection and zero-size dates of the moving and expanding jet knots shown in Figure~\ref{fig:BM538B Image and Median Model}. We derive the dates using the medians of the marginal posterior distributions as the best-fit parameters, and the 16th and 84th percentiles as the uncertainties.}
        \tablehead{
        \colhead{Jet Knot} & \colhead{Ejection Date (MJD)} & \colhead{Zero-Size Date (MJD)}
        }
        \startdata
        Knot 1 & $60206.41_{-0.07}^{+0.06}$   & $60206.78_{-0.16}^{+0.08}$  \\
        Knot 2 & $60206.36_{-0.17}^{+0.11}$   & $60206.93_{-0.06}^{+0.03}$  \\
        Knot 3 & $60206.22\pm0.03$            & $60206.5_{-0.7}^{+0.2}$     \\
        \hline
        \enddata
    \end{deluxetable}

    \subsection{Jet Ejection}

    The ejection dates of the three transient jet knots are coincident with the soft X-ray flare on 2023 September 19. This was one of the brightest X-ray flares of the outburst, although there were many other smaller flares during this period \citep[see e.g.][and the first panel of Figure~\ref{fig:lc and dps}]{2024ApJ...974..303Z, 2024arXiv241006574L}. During this period, only type-C QPOs were detected, with no detections of type-B QPOs \citep{2024MNRAS.529.4624Y, 2024ApJ...974..303Z, 2024ApJ...968..106Z, 2024MNRAS.531.1149N}. The presence of type-B QPOs are a defining feature of the soft-intermediate state \citep{2016ASSL..440...61B}, which suggests that during this period the source never transitioned from the hard-intermediate state to the soft-intermediate state, despite the ejection of transient jet knots. This may imply that the presence of type-B QPOs is more closely associated with the accretion inflow and not the jet, and that their emergence may not be directly causally linked to jet ejection. This might explain why previous associations between jet ejection and the emergence of type-B QPOs did not show a consistent sequence of events between the changes in the X-ray timing properties and the ejection of jets \citep{2009MNRAS.396.1370F, 2012MNRAS.421..468M, 2020ApJ...891L..29H, 2021MNRAS.505.3393W}.
    
    Detailed studies of the QPO evolution during the flaring period suggested that the origin of the QPOs in Swift J1727.8-1613 is Lense-Thirring precession of the inner disk \citep{2024arXiv241006574L, 2024ApJ...973...59S}, while other observations earlier in the outburst suggested the origin of the QPOs to be the precession of the jet base \citep{2024ApJ...970L..33Y}. \citet{2024arXiv241006574L} modelled a number of NICER observations of an X-ray flare a few days prior to the 2023 September 19th flare to study the evolution of the inflows. They suggested that during the flare the soft X-rays were enhanced as the inner disk rapidly extended inwards. Around this time, the corona was suppressed, contracting vertically or being ejected, and the hard X-rays were suppressed. There were no high angular resolution observations that were able to resolve any transient jet ejecta launched during this period. We speculate that a similar evolution may have occurred during the flare on 2023 September 19, with the ejection of the three transient jets occurring as the inner disk extended inwards and the corona was suppressed, with the corona then recovering as the inner accretion disk radius increased back to a similar radius to the beginning of the flare. Detailed modelling of the HXMT observations is beyond the scope of this letter. Given that this period of the outburst included multiple bright X-ray flares, these three jet knots are unlikely to be the only transient knots launched during the outburst. 

     It is rare to observe both the compact core and transient jet knots within the same observation. This may imply that the core quenched when the transient jet knots were launched and it has now re-established in the $\sim20$ hours between their ejection and this observation. Hughes et al. (submitted) present total flux density light curves that show that around the time of ejection of the three jet knots, the overall flux density of the source decreased by almost an order of magnitude at 10\,GHz, and the spectral index went from being flat to $\sim-1$, which suggests significant (if not complete) quenching of the compact jet prior to the transient jet launching. If the compact jet quenched prior to the X-ray flare \citep[e.g.][]{2020NatAs...4..697B, 2020ApJ...891L..29H}, it may have begun to re-establish shortly after the end of the flare when the QPO frequency decreased and the X-ray spectrum re-hardened. We note that in V404 Cygni, multiple transient jet knots were launched in a short period of time while the core remained unquenched \citep{2019Natur.569..374M}.

    It is not clear what processes led to the ejection of three distinct jet knots, or even if these knots were ejected at three distinct times. If knot 3 underwent deceleration before it reached the separation at which we observed it, its inferred ejection date would have been later. In this scenario the jet ejection could have coincided with the ejection of the other two jet knots, perhaps in some single ejection event around the time of the peak of the flare. The deceleration of transient jet knots due to their interaction with the interstellar medium has been seen at mas scales \citep[e.g.][]{2010MNRAS.409L..64Y, 2011MNRAS.415..306M, 2017MNRAS.468.2788R}, and it cannot be ruled out here. We were unable to constrain any intra-observational deceleration of any of the jet knots. The position angles of the three jets knots are significantly different, with a difference of $2\pm0.2\degree$ for knots 1 and 3. Even if they were ejected at a similar time, they may have not collided and interacted with each other close to their launch site. They may have been ejected at separate distinct times, although there was no obvious repeated signature of ejection of three knots in the X-ray light curves and power spectra. 
    
    The ejection of multiple distinct jet knots within a single short period of time has been observed before in low-mass X-ray binary outbursts. In MAXI J1820+070, two distinct jet knots travelling at drastically different speeds were ejected within a short period around the time of a bright, soft X-ray flare and a sudden change in the X-ray timing properties \citep{2020ApJ...891L..29H, 2021MNRAS.505.3393W}. The MAXI J1820+070 ejecta were launched as the system transitioned from the hard-intermediate to the soft-intermediate states, a transition that did not occur until much later in the outburst of Swift J1727.8-1613 \citep{2023ATel16271....1M, 2023ATel16273....1B}. In a more extreme example, during a bright flaring period of the 2015 outburst of V404 Cygni, 12 distinct jet knots were observed to have been ejected during a single 4-hour period \citep{2019Natur.569..374M}, although the ejection of these jet knots could not be associated with a clear X-ray signature due to obscuration of the inner accretion disk \citep{2017MNRAS.471.1797M}. 

    Although the precise causal connection between the changes in the accretion inflow and the ejection of these three jet knots is still unclear, this observation represents one of the most precise associations between the ejection of transient jet knots and signatures of drastic changes in the accretion inflow, with an ejection date uncertainty of $\pm40$ minutes for knot 3. The accretion/ejection association established in MAXI J1820+070 was the result of monitoring of the trajectory of the transient jet knots from mas to arcsecond scales with multiple instruments over hundreds of days \citep{2020NatAs...4..697B}, which yielded an ejection date uncertainty of $\pm30$ minutes \citep{2021MNRAS.505.3393W}. We have been able to make similarly precise measurements of the ejection dates of multiple jet knots from a single high resolution VLBI observation through the use of time-dependent self-calibrated visibility model fitting. This is a reduction by three orders of magnitude in the length of monitoring required to measure transient jet ejection dates. High angular resolution observations of transient jet launching, analysed using techniques designed to capture the intrinsic variability of the jet knots, accompanied by dense, contemporaneous X-ray monitoring, are key to furthering our understanding of the accretion/ejection connection in low mass X-ray binaries. Techniques like these will be necessary to study observations of Galactic X-ray binary jets with next generation facilities like ngEHT and ngVLA. These instruments will observe with higher angular resolution, at higher frequencies, and with higher sensitivity, resulting in a drastic increase in the amount of intra-observational variability that cannot be captured with traditional analysis techniques.
    
\section*{Acknowledgements}
We respectfully acknowledge the significant contributions made to this longstanding collaboration by Tomaso Belloni, who sadly passed away during our observing campaign. His insights and wealth of knowledge are sorely missed by his colleagues.

The authors would like to thank Andrzej Zdziarski for useful discussions.

The National Radio Astronomy Observatory is a facility of the National Science Foundation operated under cooperative agreement by Associated Universities, Inc. This work made use of the Swinburne University of Technology software correlator, developed as part of the Australian Major National Research Facilities Programme and operated under licence. This work made use of the data from the Insight-HXMT mission, a project funded by the China National Space Administration (CNSA) and the Chinese Academy of Sciences (CAS), and data and/or software provided by the High Energy Astrophysics Science Archive Research Center (HEASARC), a service of the Astrophysics Science Division at NASA/GSFC. This research has made use of the MAXI data provided by RIKEN, JAXA and the MAXI team.

CMW acknowledges financial support from the Forrest Research Foundation Scholarship, the Jean-Pierre Macquart Scholarship, and the Australian Government Research Training Program Scholarship. DMR is supported by Tamkeen under the NYU Abu Dhabi Research Institute grant CASS. TDR is an INAF research fellow. AJT acknowledges the support of the Natural Sciences and Engineering Research Council of Canada (NSERC; funding reference number RGPIN-2024-04458). VT acknowledges support from the Romanian Ministry of Research, Innovation and Digitalization through the Romanian National Core Program LAPLAS VII – contract no. 30N/2023.

The authors wish to recognise and acknowledge the very significant cultural role and reverence that the summit of Maunakea has always had within the indigenous Hawaiian community. We are most fortunate to have the opportunity to conduct observations from this mountain. 

%

\facilities{NRAO, VLBA, HXMT, MAXI}


\software{AIPS \citep{1985daa..conf..195W, 2003ASSL..285..109G}, ADS (\url{https://ui.adsabs.harvard.edu/}), Arxiv (\url{https://astrogeo.org/}), Astrogeo (\url{https://astrogeo.org/}), Astropy \citep{astropy:2013, astropy:2018, astropy:2022}, CDS \citep[Simbad;][]{2000A&AS..143....9W}, Cmasher \citep{2020JOSS....5.2004V}, Corner \citep{corner}, Dynesty \citep{2020MNRAS.493.3132S}, eht-imaging \citep{2018ApJ...857...23C}, Jupyter \citep{Kluyver2016jupyter}, Matplotlib \citep{Hunter:2007}, Numpy \citep{harris2020array}, Scipy \citep{2020SciPy-NMeth}, Stingray \citep{2019ApJ...881...39H, bachettiStingrayFastModern2024}}





\bibliography{References}{}
\bibliographystyle{aasjournal}



\end{document}